\documentclass[twocolumn]{aastex631}

\shorttitle{Spectral Steepening in Diffusive Shock Acceleration}
\shortauthors{Cristofari et al.}
\graphicspath{{./}{figures/}}

\begin{document}

\title{Microphysics of diffusive shock acceleration: impact on the spectrum of accelerated particles}

\correspondingauthor{P. Cristofari}
\email{pierre.cristofari@icjlab.in2p3.fr}

\author[0000-0002-0786-7307]{Pierre Cristofari}
\affiliation{Universit\'e Paris-Saclay, CNRS/IN2P3, IJCLab, 91405 Orsay, France}

\author[0000-0003-2480-599X]{Pasquale Blasi}
\affiliation{Gran Sasso Science Institute, via F. Crispi 7--67100, L'Aquila, Italy}
\affiliation{INFN/Laboratori Nazionali del Gran Sasso, via G. Acitelli 22, Assergi (AQ), Italy}

\author[0000-0003-0939-8775]{Damiano Caprioli}
\affiliation{Department of Astronomy and Astrophysics, University of Chicago, 5640 S Ellis Ave, Chicago, IL 60637, USA}



\begin{abstract}
Diffusive shock acceleration at collisionless shocks remains the most likely process for accelerating particles in a variety of astrophysical sources. While the standard prediction for strong shocks is that the spectrum of accelerated particles is universal, $f(p)\propto p^{-4}$, numerous phenomena affect this simple conclusion. In general, the non-linear dynamical reaction of accelerated particles leads to a concave spectrum, steeper than $p^{-4}$ at momenta below a few tens of GeV/c and harder than the standard prediction at high energies. However, the non-linear effects become important in the presence of magnetic field amplification, which in turn leads to higher values of the maximum momentum $p_{max}$. It was recently discovered that the self-generated perturbations that enhance particle scattering, when advected downstream, move in the same direction as the background plasma, so that the effective compression factor at the shock decreases and the spectrum becomes steeper. We investigate the implications of the excitation of the non-resonant streaming instability on these spectral deformations, the dependence of the spectral steepening on the shock velocity and the role played by the injection momentum. 
\end{abstract}



\section{Introduction} 
\label{sec:intro}

Particles repeatedly crossing a collisionless shock front are accelerated to non-thermal energies, as first discussed by ~\citep{axford1977,krymskii1977,bell1978,blandford1978}, a mechanism that became known as Diffusive Shock Acceleration (DSA). This mechanism is thought to play a crucial role in the acceleration of Galactic Cosmic Rays (CRs) when applied to supernova remnant (SNR) shocks \citep{blandford1978}, star cluster shocks \cite[]{seo2018,morlino2021} and many other astrophysical sources where shocks are produced~\citep[see, e.g.,][for reviews on the topic]{review2013,caprioli15p,blasi2019,gabici2019}.

DSA has attracted a lot of attention mainly because of its simplicity and its weak dependence on the poorly known microphysics (e.g., scattering properties of the particles). For instance, a shock with Mach number $M$ in a medium with  adiabatic index $\gamma_g$ is expected to produce a spectrum of accelerated particles
\begin{equation}\label{eq:DSA}
    f(p)\propto p^{-{3R/(R-1)}}; \quad R=\frac{\gamma_g+1}{\gamma_g - 1 + 2/M^2},
\end{equation}
where $R$ is the shock compression ratio, independent of the diffusion coefficient. 
The latter enters the description of the acceleration process by determining the particle confinement time, which in turn fixes the maximum achievable energy. 
In the limit of strong shocks ($M\gg 1$) and for $\gamma_g=5/3$, $R\to4$ and the spectrum becomes the well known $f(p)\propto p^{-4}$. 

This basic (test-particle) version of the theory of DSA is not suitable for the description of the rich phenomenology of CRs in the Galaxy: first, the $p^{-4}$ the spectrum is energy-divergent when extended to infinite maximum momenta, and even more important, for a given value of the maximum momentum $p_{max}$, it is possible that the total energy budget may become comparable with (or even exceed) $\rho v_s^2$, the total ram pressure from which energy can be tapped ($\rho$ and $v_s$ are the density of gas upstream of the shock and the shock speed). Clearly this conclusion is unphysical in that it would violate energy conservation. By itself this is sufficient motivation to build a theory that includes the dynamical reaction of accelerated particles on the system \citep[see, e.g.,][for reviews]{drury83, jones2001, malkov2001,review2013}. 

Inclusion of this effect leads to an exacerbation of the energetic problem, in that the spectrum becomes even harder than $p^{-4}$ at high energies. 
However a non-linear theory should account for this issue by providing mechanisms of self-regulation that inhibits too high efficiencies to be achieved. In other words, when the system becomes too efficient, the acceleration process should switch off. 

The second non-linear effect that is necessary for the theory to confront observations is the self-generation of magnetic perturbations due to the excitation of streaming instability upstream. In the absence of this effect, the maximum energy of accelerated particles is too low to be of astrophysical interest \cite[]{lagage1,bell1978}. In particular, many authors discussed the excitation of the non-resonant streaming instability \cite[]{bell2004} as the chief mechanism for magnetic field amplification at SNR shocks \cite[]{vink2012}, which in turn may help reaching very high energies \cite[]{bell2013,schure2013,schure2014,cardillo2015,cristofari2020,cristofaripev2021}.

One of the major challenges that the theory of DSA faces is that of producing spectra of accelerated particles that are substantially steeper than $p^{-4}$, required by both gamma ray observations of individual SNRs \cite[]{SNR_FERMI,SNR_HESS,caprioli11} and by the standard theory of CR transport in the Galaxy \cite[]{evoli2019}. For shocks with speed $\lesssim 3000$ km/s propagating in partially ionized media the spectrum of accelerated particles may be steeper than $p^{-4}$ due to the neutral return flux, first discussed by \cite{blasiNeutrals2012}. This phenomenon was shown to shape the gamma ray spectrum of the Tycho SNR \cite[]{Morlino2016}, but it cannot be the general solution of the problem of the steep spectra.

Plasma simulations have been used to investigate particle acceleration and generation of magnetic field at strong shocks, especially hybrid ones with fluid electrons and kinetic ions \citep[e.g.,][and references therein]{caprioli2014a,caprioli2014b, caprioli2014c}.
Unprecedentedly-long hybrid simulations \citep{haggerty2020} showed that magnetic field perturbations, amplified upstream by the Bell instability and compressed at the shock, \emph{downstream} move in the same direction as the plasma, with a velocity close to the Alfv\'en speed $v_A$ calculated in the amplified magnetic field, $v_A=\delta B_2/\sqrt{4\pi \rho_2}$. 

While linear Alfv\'en waves are expected to be transmitted/reflected at the shock \citep[e.g.,][and references therein]{scholer+71, caprioli+08, caprioli+09a}, upstream non-linear magnetic fluctuations travel with the upstream speed and overshoot when crossing the shock, retaining a net motion towards downstream, even if they are rapidly slowed down to the typical speed of magnetic fluctuations, i.e., the local Alfv\'en speed.

The drift of magnetic perturbations is due to the fact that the accelerated particles move under the action of advection and diffusion. The former is not the velocity of the background plasma (as it is often assumed) but rather the speed of the scattering centers. This is usually small enough compared with the plasma speed that the difference is inconsequential. However, as was already pointed out by~\citet{bell1978}, if it happens that the velocity of the scattering centers (either upstream or downstream) becomes an appreciable fraction of the plasma speed, the spectrum of accelerated particles may be heavily affected, by becoming either steeper or harder depending on the direction of the motion of scattering centers. For the modes generated by Bell instability, that is crucial for particle acceleration, the downstream velocity of the perturbations can become an appreciable fraction of the fluid velocity downstream, as discussed by~\citet{caprioli2020}, in the direction of having waves moving away from the shock. This causes the accelerated particles to drift away from the shock faster than in the standard situation.

This phenomenon leads to the creation of a \emph{postcursor}, where an interesting phenomenology for the CR spectra unfolds, as discussed in \cite{caprioli2020}. 
In particular, the effective compression factor felt by the accelerated particles becomes $R\approx u_1/(u_2+v_A)$, substantially smaller than $u_1/u_2$, thereby implying steeper spectra of accelerated particles. 
This effect is prominent for fast shocks, in that $v_A$ can become an appreciable fraction of the downstream plasma speed $u_2$ when the magnetic field is strongly amplified \cite[]{caprioli2020}. 
This effect was also recently discussed for a variety of astrophysical contexts  by \cite{Rebecca2021}, in the context of a semi-analytical approach to non--linear DSA. 

In the present work we investigate the spectral steepening resulting from the excitation of the non-resonant instability and the formation of a postcursor. In particular we study the dependence of this effect on the efficiency of particle acceleration at SNR shocks and on the injection momentum. The latter becomes an important parameter of the problem if the magnetic field amplification is strong enough to make the spectra steeper than $p^{-5}$, at which point the energetic of accelerated particles becomes dominated by very low energy particles. We compare our predictions with observations in a selection of SNRs for which reliable measurements of the magnetic field and of the shock velocity exist. 

The article is organized as follows: 
In \S \ref{sec:spectrum} we discuss the physics of the postcursor and the recipes for magnetic field amplification. In \S \ref{sec:magneticfield} we illustrate the implications of the postcursor in terms of spectral slope and maximum energy of accelerated particles, and level of magnetization at the shock. 
In \S \ref{sec:concl} we summarize the advances implied by the discovery of the postcursor physics and the caveats to be kept in mind, as well as the problems still left open.

\section{The spectrum of accelerated particles} \label{sec:spectrum}

In this section we describe in detail the effect of the magnetic field amplification on the spectrum of accelerated particles at the shock. 

In the test--particle limit~\citep{axford1977,krymskii1977,bell1978,blandford1978}, the shock compression ratio $R$  is derived by imposing mass, momentum and energy conservation at the shock surface. 
The spectrum of accelerated particles is universal and is a power law with slope $q=3R/(R-1)$ (Equation \ref{eq:DSA}), independent of the details of particle scattering in the shock region. 
The slope of the spectrum is shaped by two physical quantities, the energy gain per cycle of a particle crossing the shock on both sides, $\Delta p/p\simeq \frac{4}{3}\frac{u_1-u_2}{c}$, and the return probability from the downstream region, $P\simeq 1-4\frac{u_2}{c}$. In the limit of a strong shock, $M\gg 1$, the spectrum tends to its asymptotic $p^{-4}$ shape.  

This universality may be broken by several effects, both at the test particle level and at the non--linear level. At the test particle level, if the scattering centers move with respect to the plasma, as it is the case for Alfv\'en waves, the compression factor relevant for particle acceleration becomes $(u_1\pm v_{A,1})/(u2\pm v_{A,2})$, where the index indicates whether the quantity is calculated upstream (1) or downstream (2). The $\pm$ sign identifies whether the waves move in the same direction (+) or in the opposite direction (-) with respect to the plasma \cite[]{bell1978}. This effect leads to harder or softer spectra of accelerated particles, depending on the relative motion of the waves and the plasma. If the Alfv\'en speed is calculated with respect to the background magnetic field, typically this spectral deformation is very weak, unless the shock itself is weak, namely $M\sim$ a few. 

At non-linear level, as mentioned above, the test particle prediction is changed by both the dynamical action of the CRs \citep[see][for a review]{malkov2001} and by magnetic field amplification induced by accelerated particles \citep[]{haggerty2020,caprioli2020}. This latter effect is due to the fact that the magnetic perturbations generated by the CR themselves upstream of the shock, once advected downstream, move in the same direction as the plasma, at roughly the Alfv\'en speed in the amplified field \cite[]{haggerty2020}, which in turn can be an appreciable fraction of the plasma speed. 
In these conditions the spectrum of accelerated particles becomes steeper that predicted based on the test particle theory \cite[]{caprioli2020}. 

In the following, we quantify the importance of this effect while retaining a simple semi-analytical description of the phenomenon. 
We work under the assumption that at the shock, the differential spectrum of particles accelerated through DSA remains a power--law $f(p) \propto p^{-\alpha}$, thereby neglecting the curvature 
that may arise if there is a strong precursor and particles of different $p$ feel different compression ratios. 
Note that this does not necessarily imply the CR backreaction to be negligible: the postcursor is a non-linear feature that should apply to CRs of any momentum.

Following \cite{haggerty2020,caprioli2020} we  retain information about the non--linear shock modification by introducing a total plasma compression factor $R_{\rm tot}\equiv u_0/u_2$, that in the presence of CRs is larger than 4 (here $u_0$ is the plasma speed at upstream infinity). We also introduce $R_{\rm sub}\equiv u_1/u_2$ as the compression factor between immediately upstream and downstream of the shock;
for small to moderate acceleration efficiencies, we can assume $R_{\rm sub}\approx 4$.

If the CR spectrum is taken as $f(p) = A \left( \frac{p}{mc}\right)^{-\alpha}$, the  normalization constant $A$ is computed by imposing that at the shock, a fraction $\xi_{\rm CR}$ of the ram pressure is converted into CRs, which leads to 
\begin{equation}
    A= \frac{3}{4\pi} \frac{\xi_{\rm CR} \rho v_{\rm sh}^2}{m^4 c^5 I(\alpha)};
   \quad
    I(\alpha)\equiv \int_{x_{\rm inj}}^{x_{\rm max}} \text{d}x\frac{  x^{4-\alpha}}{\sqrt{1+x^2}},
\end{equation}
where we posed $x=p/m_pc$.

As discussed by \cite{caprioli2020}, the slope of the spectrum is determined by the downstream Alfv\'en speed, $v_{A,2}=R_{\rm tot}\delta B_2/\sqrt{4\pi\rho}$, where we assume that the magnetic field upstream is amplified through the excitation of the non-resonant Bell instability \cite[]{bell2004}.
The latter statement requires some additional comments, reflecting the discussion in \cite[]{cristofari2021}: the magnetic field upstream that is usually adopted to calculate the maximum energy is the one at upstream infinity, as produced by the escaping particles. This is because the current inducing the instability is the current of escaping particles far upstream of the shock. If, in first approximation, the spectrum is close to $p^{-4}$, then one can estimate the magnetic energy density as 
\begin{equation}
\frac{B_1^{2}}{4\pi}\approx 3\frac{v_{\rm sh}}{c}\frac{\xi_{\rm CR}\rho v_{\rm sh}^{2}}{\ln\left(\frac{p_{\rm max}}{m_p c}\right)},
\label{eq:saturaA}
\end{equation}
while the corresponding downstream magnetic field after compression of the perpendicular components at the shock is
\begin{equation}
\label{eq:B2Bell}
B_2=\sqrt{\frac{(1+ 2 R_{\rm tot}^2)}{3}} B_1.
\end{equation}

On the other hand, the current immediately upstream of the shock can be larger in that all particles contribute to the CR current in that region. If this is the case, then the magnetic field $B_1$ can be written as \cite[]{cristofari2020,cristofari2021}:
\begin{equation}
\label{eq:B1}
B_1=\sqrt{ 12 \pi p_{\rm inj}^{4-\alpha} \left(\frac{v_{\rm sh}}{c} \right)^{5-\alpha}    \frac{\xi_{\rm CR} \rho v_{\rm sh}^2}{(\alpha -3) I(\alpha)} },
\label{eq:saturaB}
\end{equation}
and the downstream field is 
\begin{equation}
\label{eq:B2}
B_2=\sqrt{\frac{(1+ 2 R_{\rm sub}^2)}{3}} B_1,
\end{equation}

For $\alpha=4$ one recovers the fact that the amplified field is independent of the value of $p_{\rm inj}$, a fact that remains approximately true also for $\alpha\gtrsim 4$, unless $\alpha$ approaches the critical value of $5$, for which the energy density in the form of accelerated particles becomes dominated by the particles close to the injection momentum. Notice that in our approach we try to retain as much as possible trace of the non--linear effects involved in the acceleration process, and in general $p_{\rm inj}$ is a multiple of the thermal momentum of the plasma particles downstream of the shock, $p_{\rm th}= \sqrt{2 m k_{\rm B}T_2}$. The temperature $T_2$ can be inferred from momentum conservation and reads:
\begin{equation}
    T_2 \approx \frac{m v_{\rm sh}^2}{k_{\rm B}R_{\rm tot}} \left( 1- \frac{1}{R_{\rm tot}}- \xi_{\rm B}- \xi_{\rm CR}\right),
    \label{eq:T2}
\end{equation}
that illustrates how for a shock that accelerates particles efficiently the temperature of the downstream gas drops to lower values. 
Following the parametrization of \citep{blasi2005}, which remains a good scaling even if CRs are injected into DSA via specular reflection rather than thermal leakage \citep[see][]{caprioli2015}, we assume that $p_{\rm inj}=\chi p_{\rm th}$, with $\chi=2\div 10$. 

When the shock becomes efficient in accelerating CRs, $p_{\rm th}$ decreases, thereby increasing the CR current and, in turn, the downstream magnetic field.
The net result is that the downstream Alfv\'en speed becomes larger, which eventually leads to steeper CR spectra, which is the effect that we are after.

For the calculation of the total compression factor $R_{\rm tot}$ we follow the approach of \cite{haggerty2020} (see their Appendix B1), where mass, momentum and energy conservation equations were solved together. 
Since, when spectra are very steep, non-relativistic CRs may carry a sizable fraction of the total pressure, we retain the general expression for the CR adiabatic index as:
\begin{equation}
     \gamma_{\rm CR} = 1 + \frac{1}{3}\frac{\int_{x_{\rm inj}}^{x_{\rm max}} \text{d}x \;  x^{4+\alpha} /\sqrt{1+x^2} }{\int_{x_{\rm inj}}^{x_{\rm max}} \text{d}x \; x^{2+\alpha} (\sqrt{1+x^2} -1)}.
 \end{equation}
$\gamma_{\rm CR}\to 4/3$ when CRs are fully relativistic.
 
\begin{figure}
\centering
\includegraphics[width=.5\textwidth]{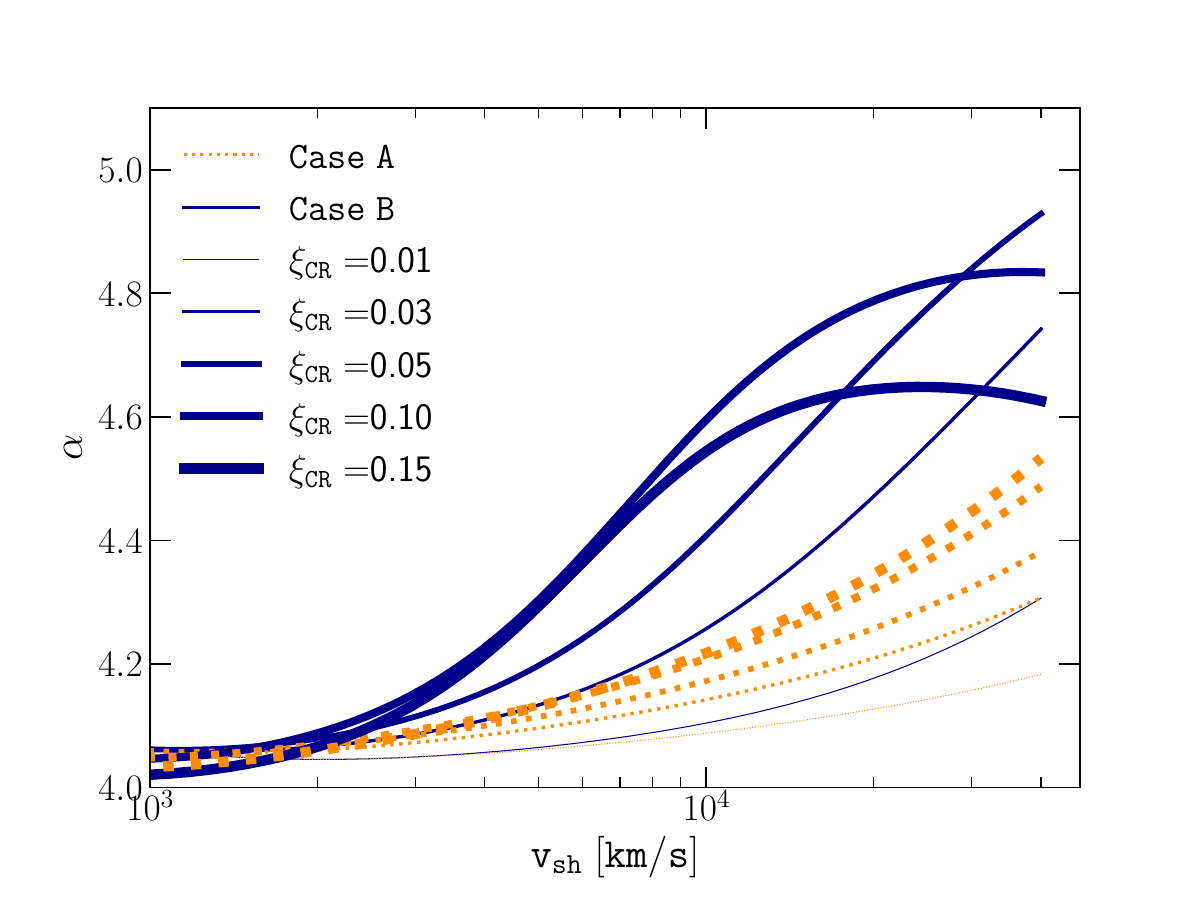}
\caption{Spectral slope $\alpha$ as a function of shock velocity $v_{\rm sh}$. 
From thin to thick lines, the CR efficiency  $\xi_{\rm CR}$ varies from 1\% to 15\%, and $p_{\rm inj}= \chi p_{\rm th}$ with $\chi=5$.
Orange dotted lines correspond to Case A, while blue solid lines correspond to Case B.}
\label{fig:alpha}
\end{figure}

\begin{figure}
\centering
\includegraphics[width=.5\textwidth]{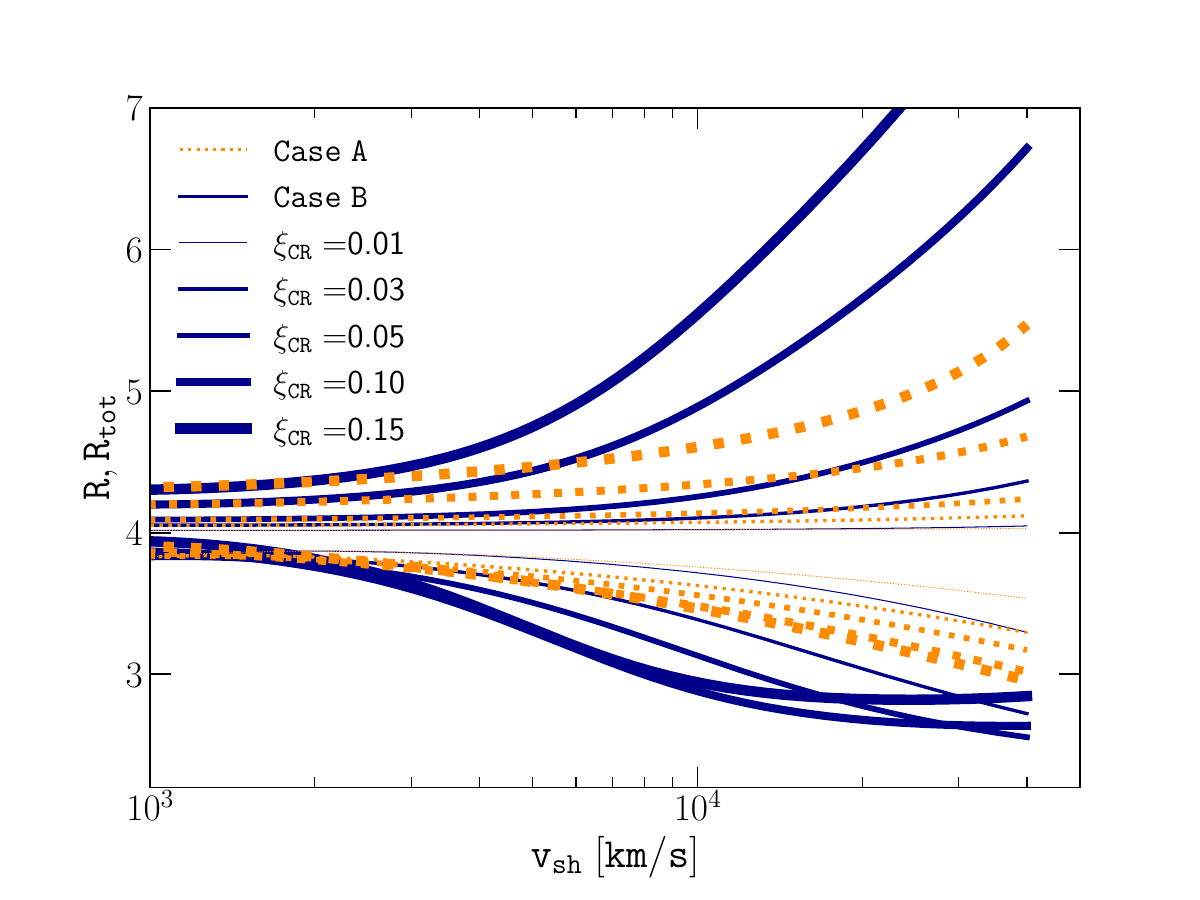}
\caption{Compression factors $R$ ($<4$) and $R_{\rm tot}$ ($\geq$4)  for the Case A (blue solid  lines) and Case B (orange dotted lines) as a function of the shock velocity $v_{\rm sh}$. 
From thin to thick lines, the CR efficiency varies $\xi_{\rm CR}$ from 1\% to 15\%, always with $\chi=5$.}
\label{fig:Rtot}
\end{figure}

This leads to a fourth-order algebraic equation for $R_{\rm tot}$. As discussed in much of the literature on non--linear theory of DSA~\citep[e.g.,][]{Ellison1999,blasi2002}, this procedure may be tricky because it requires the knowledge of the escape flux, which in turn requires the solution of the transport equation for accelerated particles \cite[]{diesing2021}. 
However, for the cases of interest here, the spectrum becomes sufficiently steeper than $p^{-4}$ that the escape flux can be neglected and the expressions derived above are good approximations to the exact ones. 

The compression factor relevant for accelerated particles (namely the compression factor of the velocities of the scattering centers) is 
 \begin{equation}
     R\approx \frac{v_{\rm sh}}{\frac{v_{\rm sh}}{R_{\rm tot}}+ f_{v_{\rm A}} v_{\rm A,2}},
     \label{eq:R}
 \end{equation}
where the numerical factor $f_{v_{\rm A}}\leq 1$ has been introduced to account for the fact that the bulk drift velocity of scattering centers might be a fraction of $v_{A,2}$, if some of the waves move in the opposite direction to the plasma. This factor is {\it a priori} set to 1 unless mentioned otherwise. 
 
The system of equations returning $R_{\rm tot}$, $R$, $\alpha=3R/(R-1)$ can be solved with an iterative procedure for a given value of the CR acceleration efficiency $\xi_{\rm CR}$ and injection momentum $\chi$, and for given shock parameters (shock velocity, Mach number).

As we show below, the spectral steepening due to magnetic field amplification depends rather strongly on the prescriptions adopted for the magnetic field downstream. In order to illustrate this effect, we consider two situations:

\begin{itemize}
    \item Case A, in which the magnetic field is produced only far upstream by escaping particles with momentum close to $p_{\rm max}$ (Eqs.~\ref{eq:saturaA} and \ref{eq:B2Bell});
    \item Case B, in which the magnetic field is amplified by the anisotropy (current) of the CRs diffusing in the precursor (Eqs.~\ref{eq:saturaB} and \ref{eq:B2}).
\end{itemize}
These two prescriptions are meant to bracket our ignorance of the actual total level of magnetic field amplification \citep[also see][for an extended discussion]{cristofari2021}:
Case A represents a lower limit on the amount of turbulence produced,  while Case B also accounts for the current in CRs at any momentum, and in fact it depends explicitly on $p_{\rm inj}$ and $\alpha$.

The spectral slope of the accelerated particles, $\alpha$, and the corresponding compression factors $R$ and $R_{\rm tot}$ are shown in Figure\ref{fig:alpha} and Figure \ref{fig:Rtot} respectively, for $\chi=5$ and different values of the acceleration efficiency between $1\%$ and $15\%$, typical of SNRs as sources of the bulk of Galactic CRs.  
The orange dotted curves in Figure \ref{fig:alpha} illustrate the results in Case A for the magnetic field, while the blue solid curves refer to Case B. 
All curves in Fig~\ref{fig:Rtot} refer to Case B, where the strongest modifications are expected (as also visible in Figure \ref{fig:alpha}). 

The spectral steepening caused by the formation of the postcursor shows the expected trend with shock velocity: increasing the shock speed the strength of the CR-induced magnetic field gets larger and as a consequence the waves' velocity downstream becomes larger and the spectrum correspondingly steeper. 
The effect is much more prominent in Case B than in Case A. 

In both cases, for the fiducial values of the parameters adopted here, the spectrum of accelerated particles remains harder than $p^{-5}$, so that the role of $p_{\rm inj}$ is never of overwhelming importance (we will comment later on this point). 
In Case B, for very large shock velocity and high CR acceleration efficiency, this conclusion may eventually become invalid, but these situations appear of rather limited physical interest. 
On the other hand, spectra close to $p^{-5}$ are in fact retrieved for the so-called radio supernovae, a finding consistent with observations \cite[see, e.g.,][]{ChevalierRadio,Soderberg2010,Kamble2016}.

A quick inspection of Fig~\ref{fig:Rtot} reveals that although the overall spectrum becomes steeper for higher shock speed, reflecting a lower compression ratio of the scattering centers $R$, the total compression factor $R_{\rm tot}$ remains larger than 4 and in fact increases for larger shock speeds and for high CR acceleration efficiencies. 
Gas compression ratios of order 6-7 have been reported in X-ray observations of single SNRs \citep[e.g.,][]{warren+05, gamil+08}.

Moreover, the fact that the spectrum of accelerated particles becomes appreciably steeper than $p^{-4}$ as a consequence of the formation of a postcursor provides support to the idea that the magnetic field immediately upstream of the shock should be estimated using Equation \ref{eq:B1} (Case B), leading to a somewhat larger estimate for such field compared with the standard expression based on Bell instability. It is however important to understand that the two are compatible with each other: the Bell recipe refers to the magnetic field at upstream infinity, where only escaping particles can reach, while Equation \ref{eq:B1} applies to the magnetic field immediately upstream, generated through the same process by all CRs with $p>p_{\rm inj}$.
While bearing in mind that the actual saturation may depend also on complex wave-wave interactions (e.g., direct and inverse cascades), damping, and turbulent amplification in the self-generated density fluctuations \citep[e.g.,][]{reville+12,caprioli+13}, in the rest of this article we will adopt Case B as our reference scenario.

\section{Impact of microphysics on spectral slope, magnetization, and CR maximum energy} \label{sec:magneticfield}

The non-linear effects that ensue from efficient DSA on one hand foster particle acceleration to high energies, but on the other hand make it a mechanism that is strongly dependent upon phenomena occurring on small scales, which are hard to access observationally. 
In this sense, the spectral steepening discussed above relies on physical intuition and the results of numerical simulations \cite[]{haggerty2020,caprioli2020}. 
In this section we describe the dependence of the main outcomes of DSA (CR spectral slope, maximum CR energy, magnetic field amplification) on the less known parameters of the theory. 

\subsection{CR spectral slope}
For selected cases, the theoretical framework described above returns meaningful results: for a Tycho-like SNR, where the shock velocity is $\sim 5000$ km/s and an acceleration efficiency of $\sim 10\%$ is required, one would expect a spectrum with a slope $\sim 4.3$, based on Figure \ref{fig:alpha}, which compares well with the slope inferred from a multi-wavelength analysis \cite[e.g.,][]
{morlino2012,slane2014}. 
For fast shocks associated with the so-called radio supernovae \cite[]{ChevalierRadio,Soderberg2010,Kamble2016}, observations require a spectral slope close to $\lesssim 5$, in good agreement with our prediction (see Figure \ref{fig:alpha} for shock velocity  $\gtrsim 3\times 10^4$ km/s and $\xi_{\rm cr}\sim 5\%$). 
These findings are consistent with those recently put forward by \citet{Rebecca2021}, who used a different prescription for the saturation of the magnetic field, which is similar to Case B but without accounting for the possible dependence on $\alpha$.
In general, we can see that varying the CR acceleration efficiency between a few and 15\% does not have a large impact on the predicted spectral slope; 
for Tycho-like parameters, for instance, we find $4.1\lesssim \alpha\lesssim 4.3$, a variation of 5\% only when changing $\xi_{\rm cr}$ by a factor of 5.
We have also tried changing $\chi$ between 2 and 10, and results do not change appreciably, with $4\lesssim \alpha\lesssim 4.4$ for Tycho.
In this respect, it is worth noticing that:
1) varying $\chi$ between 2 and 10 corresponds to varying the fraction of injected CRs by several orders of magnitude, since we are moving on the exponential tail of the Maxwellian \cite[e.g.,][]{blasi2005};
2) in a kinetic approach, $\chi$ and $\xi_{\rm cr}$ are not independent, and an increase in the former implies a decrease of the latter \citep[e.g.,][]{diesing2021};
3) PIC simulations suggest that the most likely values of $\chi$ are in the range 3-4 \citep{caprioli2014a,caprioli2015,park+15}.

\subsection{Shock magnetization}
\begin{figure}
\centering
\includegraphics[width=.5\textwidth]{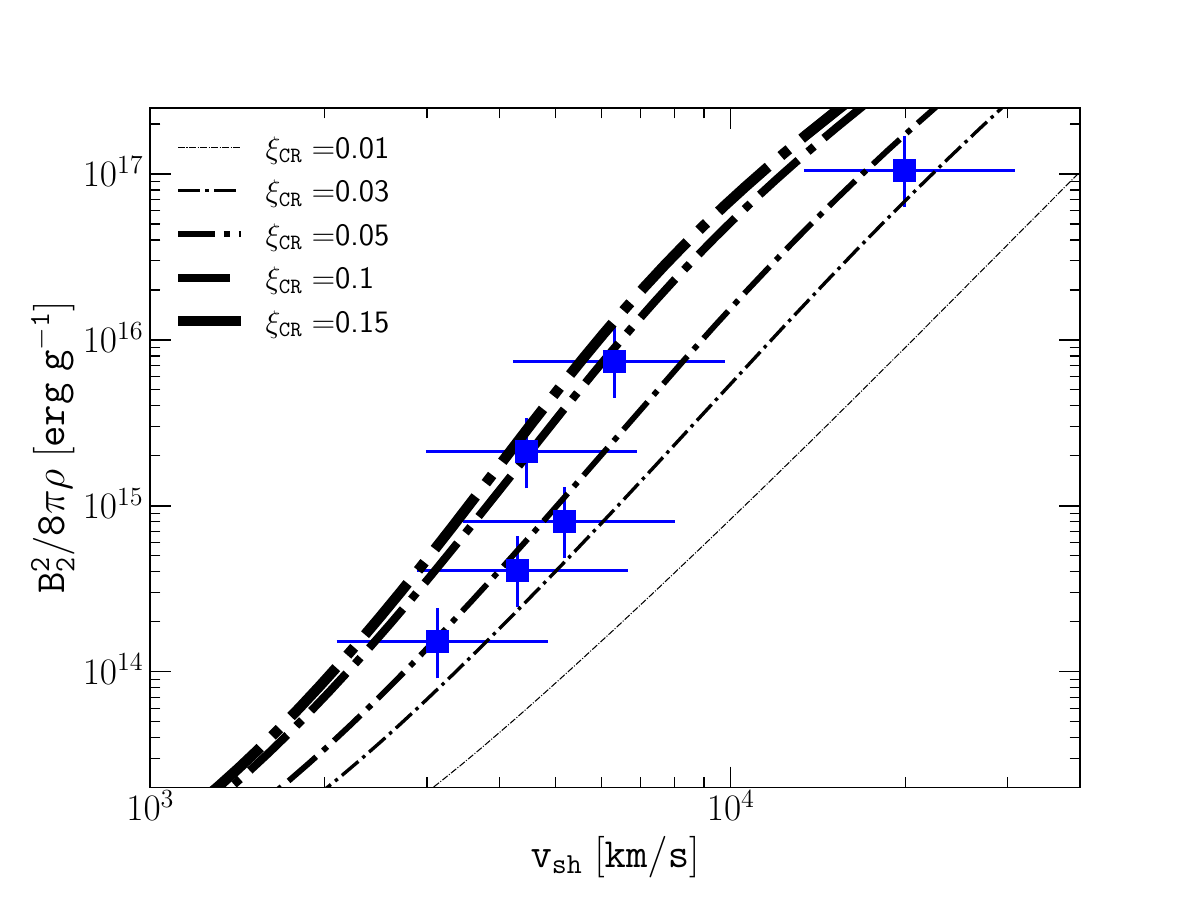}
\caption{Downstream magnetization as a function of shock velocity $v_{\rm sh}$. 
The black dot--dashed lines correspond to the values of $\alpha$ calculated for CR efficiency $\xi_{\rm CR}$ between  1\% and 15\%. }
\label{fig:vink3}
\end{figure}

In Figure \ref{fig:vink3} we show the level of magnetization of the downstream plasma for SNR shocks moving with different velocities as reported by \cite{vink2012}. 
The curves show the results of our calculations for the self-generated shock magnetization for different values of the CR acceleration efficiency, as indicated in the legend. 
One can immediately see that the standard $B^2\propto v_s^3$ scaling, typical of the saturation of the non-resonant instability \citep{bell2004}, is modified by the presence of the postcursor. 
In general, the data points seem to be better described when the postcursor physics is considered and the acceleration efficiency is around 5-10\%. 
For shock velocities $\lesssim 2000$ km/s the resonant instability becomes dominant and the magnetic field downstream becomes appreciably smaller, although the trend of $B^2$ with the shock speed becomes milder.  
Also in this case data points are in good agreement with CR acceleration efficiencies between 3 and 15\%, with the curves in Figure \ref{fig:vink3} that tend to become asymptotically close for increasingly larger efficiencies.

 \subsection{CR maximum momentum}
\begin{figure}
\centering
\includegraphics[width=.5\textwidth]{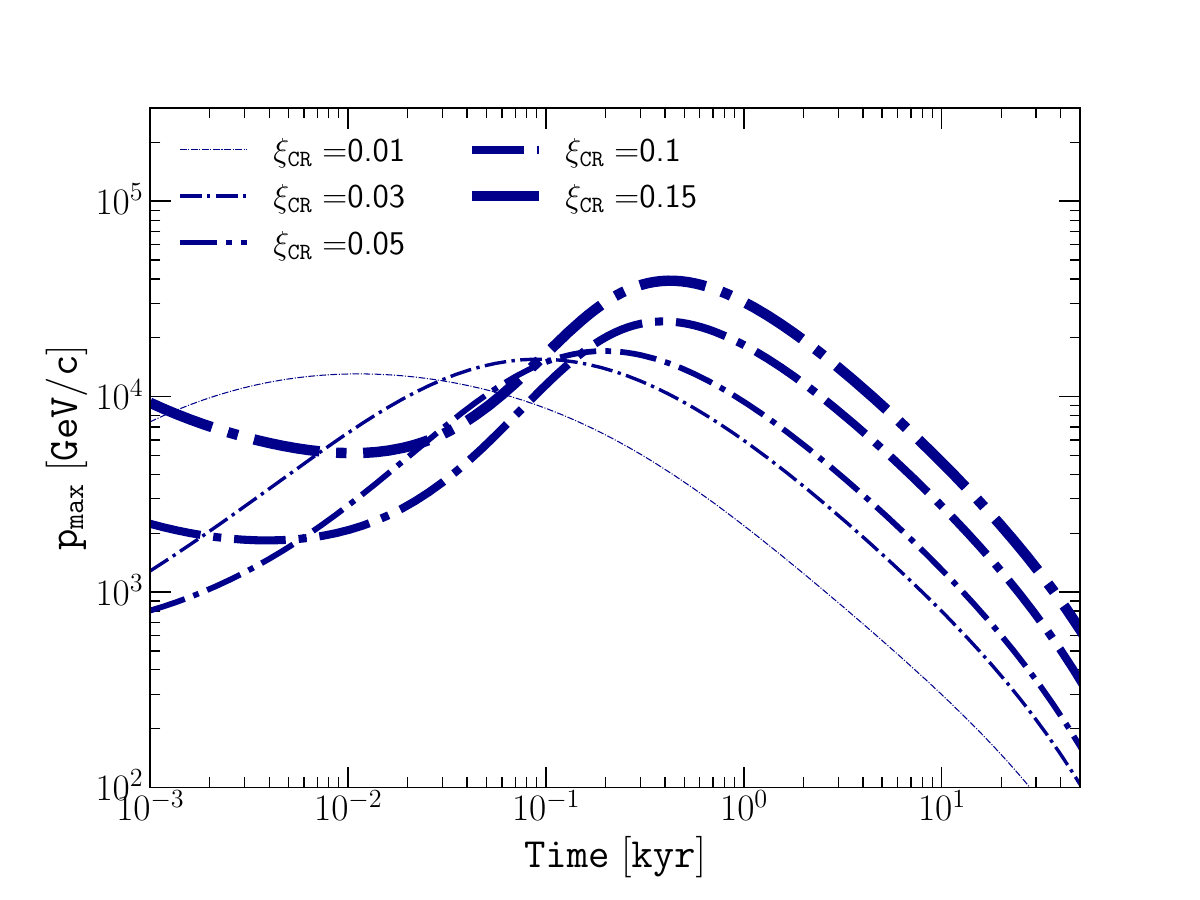}
\caption{Temporal evolution of the maximum energy of accelerated particles for a SNR from a typical type Ia SN expanding in a uniform ISM of density $n=1$ cm$^{-3}$. The CR efficiency varies from 1\% to 15\% (thin to thick).}
\label{fig:pmax}
\end{figure}

For fast SNR shocks, the maximum energy can be estimated by requiring that the growth rate $\gamma_{\rm max}$ associated to wavelength $k_{\rm max}$ reaches saturation after a few (say, $\approx 5$) $e$-folds \citep{bell2013,schure2013,cardillo2015,cristofari2020}:
\begin{equation}
    \int_0^{t} \text{d}t' \gamma_{\rm max}(t') \approx 5
\end{equation}
This leads to a maximum momentum of protons that generally depends on $\alpha$ and reads: 
\begin{equation}\label{eq:pmax}
\left(\frac{p_{\rm max}}{mc} \right)^{\alpha-3} = \frac{3e r_{\rm sh}}{10 m c^2}\frac{\sqrt{4 \pi \rho}}{c} \frac{\xi_{\rm CR} v_{\rm sh}^2}{(\alpha-3)I(\alpha)},
\end{equation}
where $r_{\rm sh}$ is the SNR shock radius. 
In the case of a remnant from a typical thermonuclear SN explosion (type Ia SN), expanding in a uniform ISM, the temporal evolution of the shock radius and velocity are well described by self--similar solutions~\citep{chevalier1982,tang2017}. 
The corresponding estimated value of $p_{\rm max}$ is shown in Figure \ref{fig:pmax}.

Figure \ref{fig:pmax} shows how the spectral steepening induced by the postcursor leads to a drastic suppression of the current of escaping CRs at early times (large shock velocity), so that $p_{\rm max}$ is reduced and in fact increases with time instead of decreasing, at least for the first few hundred years of evolution of the SNR.
As expected, the effect is more prominent when the acceleration efficiency is higher. 

When the postcursor is neglected \citep[see, for instance,][]{cristofari2020}, $p_{\rm max}(t)$ generally decreases with time: the highest maximum energy is reached at very early times, when however the mass processed by the shock is small.
The actual position of the high-energy cut-off in the overall CR spectrum released by a SNR is hence due to a trade off between achieving a large $p_{\rm max}$ and processing large amount of mass.
Interestingly, during the entire evolution of our fiducial type-Ia SNR, $p_{\rm max}$ remains below $\lesssim 50$ TeV, rather at odds with the results of current gamma--ray observations and a factor of a few smaller than what expected in type Ia SNRs for a $p^{-4}$ CR spectrum \citep[e.g.,][]{bell+11,cristofari2021}. 
This finding highlights how the current understanding of the generation of magnetic turbulence in the shock precursors (Equations \ref{eq:saturaA} and \ref{eq:saturaB}) may be incomplete, especially when steep spectra are involved.

Below we consider the dependence of the spectral steepening on two microphysical parameters of our problem, namely the CR injection momentum, and the fraction of the downstream Alfv\'en speed that enters the transport equation. 

\begin{figure}
\includegraphics[width=.5\textwidth]{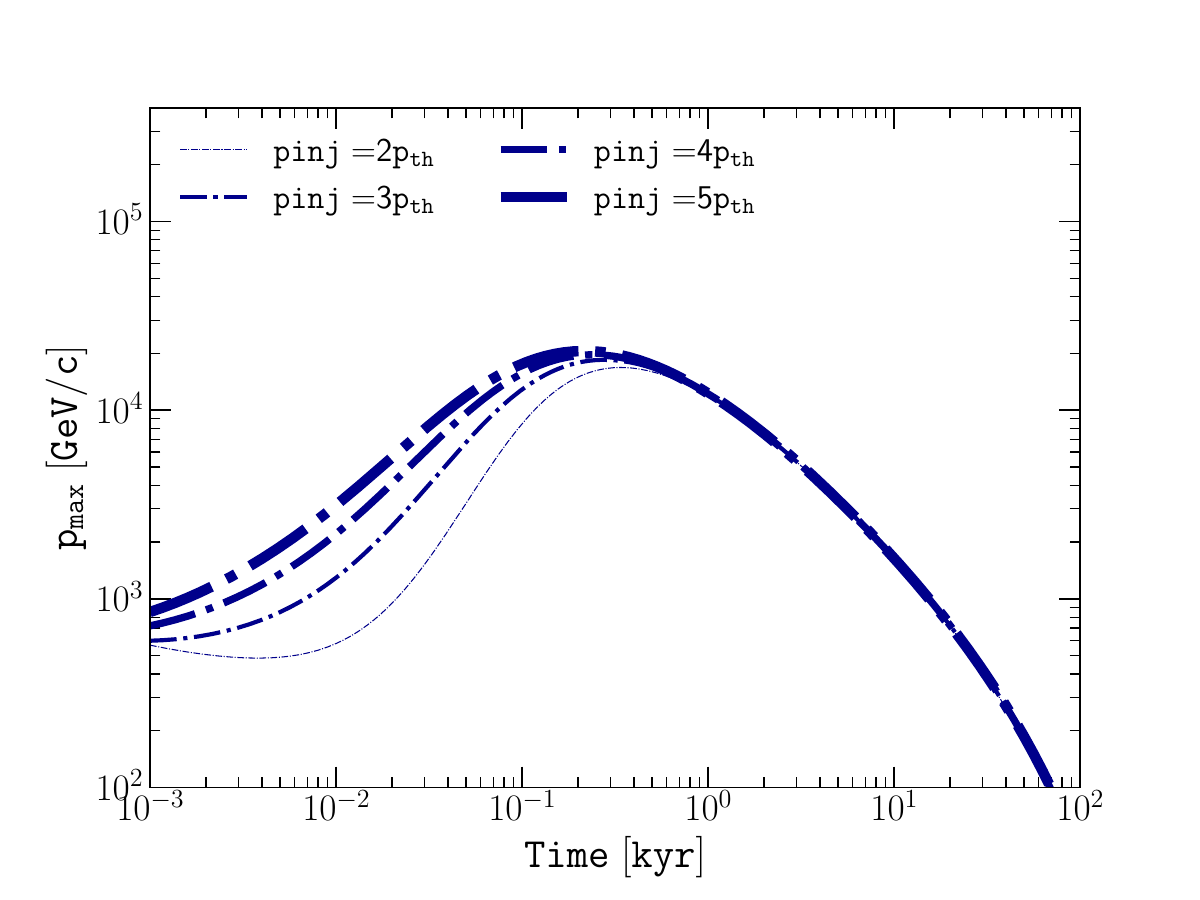}
\caption{From thin to thick lines, $p_{\rm inj}$ varies from 1 to 5 times $v_{\rm sh}/c$.  The CR efficiency is $\xi_{\rm CR}=$5\%. 
Time evolution of the maximum momentum of accelerated particles for a type Ia SNR expanding in uniform ISM of density $n=1$ cm$^{-3}$.}
\label{fig:alpha_pinj}
\end{figure}

Changing the injection  momentum does not drastically change the conclusions illustrated above about $p_{\rm max}$, as shown in Figure \ref{fig:alpha_pinj}, where $p_{\rm inj}$ is scaled as a function of the shock velocity $v_{\rm sh}$ \citep[see, e.g.,][]{caprioli2014a,caprioli2015}, rather than of the post-shock thermal speed.
Increasing the value of $p_{\rm inj}$ reduces the CR current immediately upstream of the shock, and hence the magnetic field at the same location. 
The compressed magnetic field is also lower, which results in a lessened spectral steepening, a larger current in the form of escaping protons and hence a larger value of the maximum momentum. 
However, since $\alpha$ remains appreciably smaller than 5, the changes in $p_{\rm max}$ induced by a change of $p_{\rm inj}$ are negligible. 

When spectra are steeper than $p^{-4}$, most of the magnetic field amplification occurs in the precursor rather than because of escaping particles, and that field, while contributing to the postcursor physics, does not help achieving larger values of $p_{\rm max}$.
Eventually, the effect on the total CR spectrum released by a SNR is a global steepening at $p\ll p_{\rm max}$ and a suppression by a factor of a few in the expected value of $p_{\rm max}$. At $p>p_{\rm max}$, the effect of the postcursor is only to affect the extent to which the spectrum departs from an exponential, since that part is contributed by early times when the shock velocity is larger, the instantaneous spectrum is expected to be steeper, and small amounts of mass are processed (lower normalization of the spectrum of escaping particles), as discussed by \citet{cristofari2021}.

In any case, the steep spectra observed in radio SNe and in young SNRs worsen, rather than alleviate, the problem of achieving multi-PeV energies in most SNRs.

\subsection{Parametrizing the postcursor strength}

\begin{figure}
\centering
\includegraphics[width=.5\textwidth]{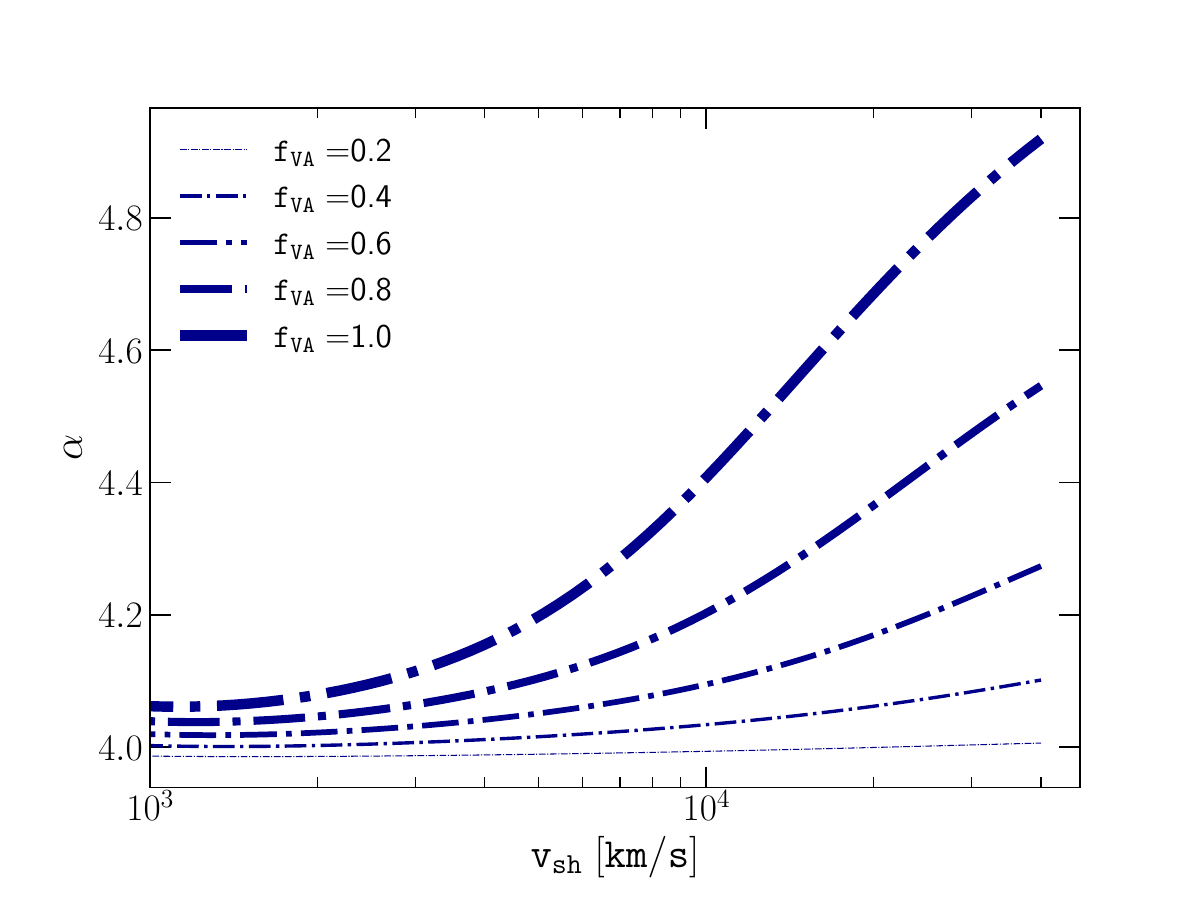}
\centering
\includegraphics[width=.5\textwidth]{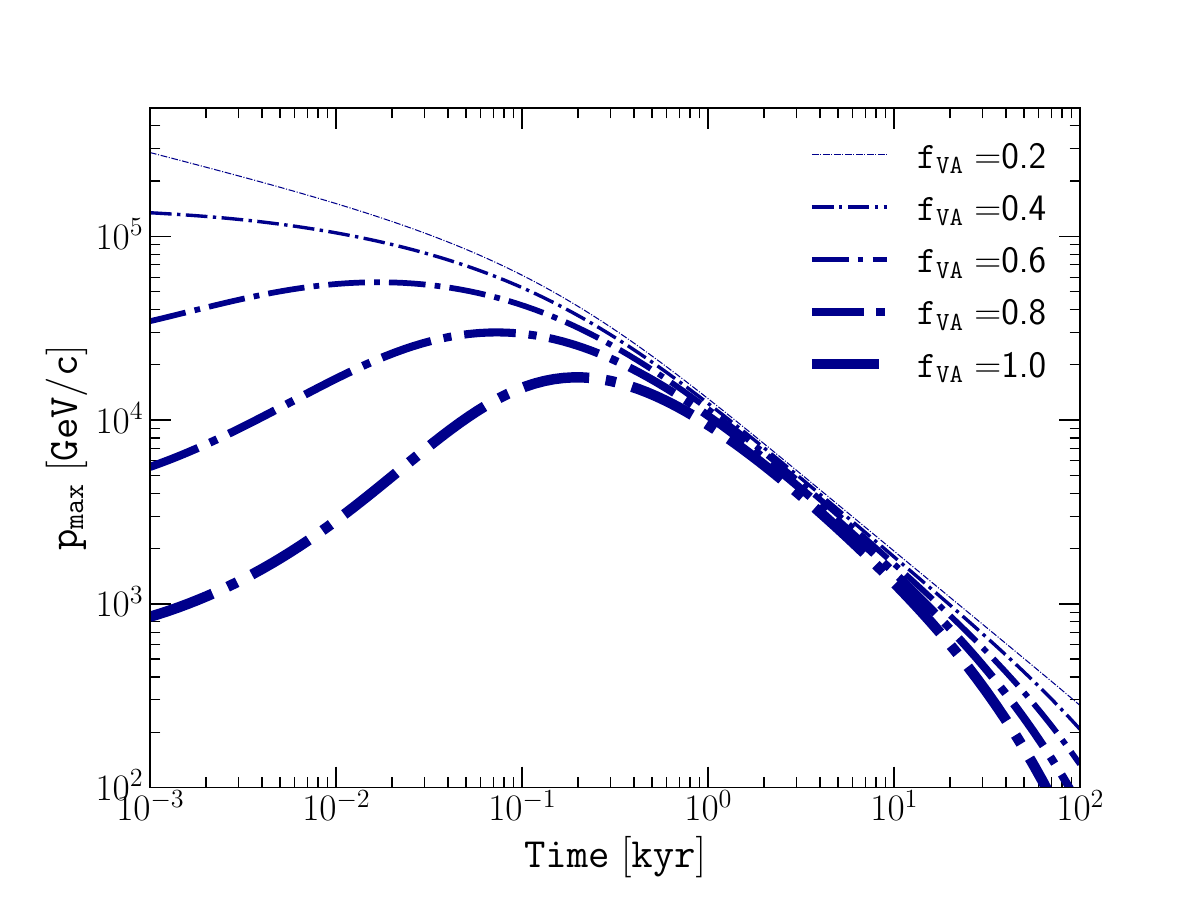}
\caption{From thin to thick lines, $f_{v_{\rm A}}=0.2,0.4,0.6,0.8$ and 1. Top panel: Spectral slope $\alpha$ vs. shock velocity $v_{\rm sh}$. Bottom panel: Time evolution of the maximum energy of accelerated particles for a SNR from a typical type Ia SN expanding in uniform ISM of density $n=1$ cm$^{-3}$. The CR efficiency is $\xi_{\rm CR}=$5\%, and $p_{\rm inj}$ is taken as in Equation~\ref{eq:T2}.}
\label{fig:alpha_fVA}
\end{figure}

The simulations by \citet{haggerty2020} and \citet{caprioli2020} suggested that the relative drift between CRs and thermal plasma in the postcursor is of order of the local Alfv\'en speed, but pinpointing its exact value for arbitrary shock velocities is challenging. We parametrize our ignorance by varying the parameter $f_{v_{\rm A}}$, as shown in Figure \ref{fig:alpha_fVA}, where a CR acceleration efficiency $\xi_{\rm CR}=5\%$ was adopted. 
In the top panel we show the spectral slope as a function of shock velocity and in the bottom panel the maximum momentum as a function of time. 
The spectral steepening due to the postcursor is strongly reduced when $f_{v_{\rm A}}$ is reduced even by a small amount; as a direct consequence, $p_{\rm max}$ correspondingly increases. 
In the limit $f_{v_{\rm A}}\to 0$, the standard results are recovered and $p_{\rm max}$ becomes again a monotonically decreasing function of time. 

\section{Conclusions}
\label{sec:concl}

The test-particle theory of DSA has well defined predictions that appear to be independent of the poorly known microphysical aspects that characterize the transport of particles in the shock region. 
In the context of the test-particle theory, the spectrum of accelerated particles is a pure power law with a slope solely determined by the shock Mach number, $f(p)\propto p^{-4}$ for any strong shock. 

The simplicity of the predictions of DSA in the test-particle assumptions is lost when the non-linear effects are included; 
such effects are the very reason why the theory is interesting in the first place, in that in their absence the maximum energy is too small to be of astrophysical interest.
Nevertheless, the theory develops several inconsistencies due to the energy divergent spectrum predicted in the context of the test particle approach. 

These non-linearities manifest themselves in several different ways.
First, the dynamical action of accelerated particles due to pressure gradients in the CR distribution upstream leads to departures from a power law \cite[e.g.,][]{malkov2001}. 
The second, and perhaps, most important non-linear aspect of DSA is in the self-generation of magnetic perturbations due to the same accelerated particles \cite[]{bell1978,lagage1,amato2006,bell2004,AmatoBlasi2009}. 
This phenomenon is thought to be responsible for the substantial increase in the maximum momentum achievable by CRs, due to the more effective confinement in the shock region. 
The non-resonant instability \cite[]{bell2004} is especially important in this sense because of its large growth rate and saturation to fields much larger than the initial one. 
The magnetic field produced through the excitation of this instability is compatible with the observation of thin X-ray non-thermal filaments in virtually all young SNRs \cite[e.g.,][]{vink2012, volk2005}. 

The apparent success of this theoretical development is balanced by the failure of the theory in explaining the spectral shape of the accelerated particles, as inferred from radio observations \cite[e.g.,][]{ChevalierRadio}, gamma-ray observations \cite[e.g.,][]{caprioli11} and from measurements of the energy dependence of the secondary/primary ratios in CRs \cite[e.g.,][]{evoli2019}: 
while the theory would require spectra generally harder than $p^{-4}$, observations lead to spectra $f(p)\propto p^{-(4.2\div 4.3)}$ or even $f(p)\propto p^{-5}$ for very fast shocks in radio SNe. 

The recent hybrid simulations by \cite{haggerty2020,caprioli2020} showed that the magnetic perturbations excited by CRs upstream of the shock, once advected across the shock in the downstream region, move with a speed $\sim v_A$ with respect to the background plasma. 
When the field is sufficiently amplified, $v_A$ can become an appreciable fraction of $u_2$ and this results in a reduced effective shock compression ratio and a corresponding substantial spectral steepening. 
These effects, which manifests themselves via the formation of a shock postcursor, are investigated in this paper.
In particular, we investigate how poorly-constrained microphysical ingredients (development and saturation of the amplified magnetic field, minimum CR momentum, and exact drift speed of CRs with respect to the thermal plasma in the postcursor) may impact macroscopical observables such as the CR slope, the post-shock magnetization, and the expected CR maximum momentum.

In Figure \ref{fig:alpha} and \ref{fig:Rtot} we show that, for a reasonable range of CR acceleration efficiencies and CR injection momenta, the postcursor steepening is very prominent when the shock is very fast ($v_{sh}$ of tens of thousand km/s) and it can lead to spectra of accelerated particles close to $p^{-5}$, while generally producing spectra slightly steeper than $p^{-4}$ for historical SNRs \citep[also see][]{diesing2021}.

The strength of the postcursor effect on the spectrum is however sensitive to the value of the magnetic field immediately upstream of the shock, and it turns out to be much milder when only the standard Bell field at upstream infinity is accounted for. 
If, on the other hand, one estimates the additional field generated on small spatial scales immediately upstream due to the local CR current, following \cite{cristofari2021}, the effect becomes much more prominent. 
We compared the magnetization of the downstream region versus observations of SNRs with different shock velocity \cite[]{vink2012} and we found good agreement for acceleration efficiency of few to ten percent.
Additional effects that may change the actual level of magnetization (either extra amplification due to fluid instabilities or damping mechanisms are discussed in \citet{cristofari2021}.


Very importantly, the spectral steepening causes a reduction of the current in the form of escaping particles, the ones most responsible for the determination of $p_{\rm max}$. 
Hence the maximum momentum decreases to $\lesssim 50$ TeV, a factor of a few lower than what expected for a $p^{-4}$ spectrum, and potentially at odds with gamma-ray observations of SNRs. 
In general, steep CR spectra can produce strong precursor currents and hence high levels of shock magnetization, but hardly produce turbulence far upstream, which is necessary to achieve large $p_{\rm max}$ values.
The tension between the observed CR knee and the maximum energy produced by different classes of SNRs \citep[e.g.,][]{bell+11, cardillo2015, cristofari2020}, is only exacerbated when CR spectra are as steep as inferred from multi-wavelength observations of the same objects.

Another effect that we discuss here is associated with the dependence of the postcursor phenomenology on the details of the reflection and absorption of magnetic perturbations at the shock: 
as discussed by \cite{haggerty2020}, the velocity of these perturbations in the downstream region is of the order of the Alfv\'en speed in the amplified field, but a leap of faith is required to extrapolate the results of kinetic simulations to shocks with much larger Mach numbers, as those pertaining to SNRs.
In general, the effective speed of the magnetic irregularities in the postcursor will be a fraction of the Alfv\'en speed, so we investigated this effect by parametrizing the velocity as $f_{v_{\rm A}}v_A$, where $f_{v_{\rm A}} \leq 1$. 
In the limit $f_{v_{\rm A}}\to 0$, the effects of the postcursor disappear. 

The magnitude of the spectral steepening and the value of $p_{\rm max}$ are quite sensitive to the value of $f_{v_{\rm A}}$. 
Changing $f_{v_{\rm A}}$ from 1 to 0.4 changes the slope at, say, $v_{\rm sh}=10^4$ km/s from $\sim 4.4$ to $\sim 4.05$ and the maximum energy at the beginning of the Sedov-Taylor phase from $\sim 15$ TeV to $\sim 50$ TeV.

We have showed how uncertain aspects of the microphysics of particle acceleration at shocks, namely the saturation of CR-driven instabilities and the actual drift of magnetic fluctuations and CRs with respect to the thermal plasma in the postcursor, may have profound phenomenological implications.
Generally, non-linear effects spoil the apparent simplicity of DSA as developed in the test particle regime, with the spectral slope becoming a function of the self-generated magnetic field and its topology.

While numerical simulations help in identifying the most important aspects of the problem and the main physical mechanisms that may be at work, this may not always be sufficient, especially when the outcome requires integration upon extended periods of time of a SNR evolution, during which different conditions may be present. 
On the other hand, some observables, such as the gamma ray emission form the shock region, are most sensitive to the conditions at the time of acceleration. 
In this case, the effects of the postcursor may be easier to identify and to study in more detail.



\begin{acknowledgments}
This project has received funding from the European Union's Horizon 2020 research and innovation program
under the Marie Skłodowska-Curie grant agreement no. 945298.
\end{acknowledgments}

\bibliography{sample631}{}
\bibliographystyle{aasjournal}



\end{document}